\pdfoutput=1
\documentclass[aps,showpacs,prl,twocolumn,superscriptaddress]{revtex4-1}
\usepackage{epsfig}
\usepackage{amsmath,amssymb}
\usepackage{hyperref}
\usepackage{color}
\usepackage{colortbl}
\usepackage{placeins}
\usepackage{ragged2e}

\begin{document}
\title{\textit{Ab-initio} multi-scale simulation of high-harmonic generation in solids}
\author{Isabella Floss}\thanks{corresponding author, e-mail: isabella.floss@tuwien.ac.at}
\author{Christoph Lemell}
\author{Georg Wachter}
\author{Valerie Smejkal}
\affiliation{Institute for Theoretical Physics, Vienna University of Technology, Wiedner Hauptstr.\ 8-10, A-1040 Vienna, Austria, EU}
\author{Shunsuke A. Sato}
\affiliation{Max Planck Institute for the Structure and Dynamics of Matter, Luruper Chaussee 149, D-22761 Hamburg, Germany, EU}
\author{Xiao-Min Tong}
\author{Kazuhiro Yabana}
\affiliation{Center for Computational Sciences, University of Tsukuba, Tsukuba 305-8577, Japan}
\affiliation{Graduate School of Pure and Applied Sciences, University of Tsukuba, Tsukuba 305-8571, Japan}
\author{Joachim Burgd\"orfer}
\affiliation{Institute for Theoretical Physics, Vienna University of Technology, Wiedner Hauptstr.\ 8-10, A-1040 Vienna, Austria, EU}

\date{\today}

\begin{abstract}
High-harmonic generation by a highly non-linear interaction of infrared laser fields with matter allows for the generation of attosecond pulses in the XUV spectral regime. This process, well established for atoms, has been recently extended to the condensed phase. Remarkably well pronounced harmonics up to order $\sim 30$ have been observed for dielectrics. We present the first \textit{ab-initio} multi-scale simulation of solid-state high-harmonic generation. We find that mesoscopic effects of the extended system, in particular the realistic sampling of the entire Brillouin zone, the pulse propagation in the dense medium, and the inhomogeneous illumination of the crystal have a strong effect on the formation of clean harmonic spectra. Our results provide a novel explanation for the formation of clean harmonics and have implications for a wide range of non-linear optical processes in dense media.
\end{abstract}
\maketitle

The generation of high harmonics (HHG) in the non-linear interaction of intense ultrashort infrared (IR) laser pulses with matter has turned out to be a highly successful route towards the generation of attosecond pulses in the EUV and XUV spectral regimes \cite{Drescher2001,Zhao2012,Takahashi2013,Hammond2016}. It has become the workhorse of investigation of a vast array of electronic processes on the attosecond time scale \cite{Krausz2009}. Expanding the range of accessible photon energies and intensities faces, however, fundamental limitations. Experimental and theoretical investigations have established a scaling of the cut-off energy $E_\mathrm{cut}\propto\lambda^2$ for HHG from atoms in the gas phase raising hopes to reach ever higher photon energies by increasing the wavelength $\lambda$ of the driving laser pulse. However, the intensity in the cut-off region was found to scale unfavorably $I_\mathrm{cut}\propto\lambda^{-5.3}$ due to the large spatial dispersion of the electron wave packet upon return to its parent atom \cite{Tate07,Schiessl07,Colo08,Fro08,Schiessl}. Propagation effects in gas filled capillaries have been found to partially offset this suppression at high $\lambda$ \cite{Popmintchev2012}.

Extending HHG to the condensed phase promises to overcome some of these limitations to enable compact and brighter light sources and to open up the novel field of solid-state photonics on the attosecond scale. The recent observation of HHG in solids for intensities below the damage threshold \cite{Ghim2011,Muecke2011,Ghimire2014,Schubert2014,Hohenleutner2015,Luu2015,Ndabashimiye2016} suggests opportunities for controlling electronic dynamics \cite{Hohenleutner2015,Luu2015} and for an all-optical reconstruction of the band structure \cite{Vampa2015b}.

The observed solid-state HHG substantially differs from the corresponding atomic spectra. For example, while for atoms the cut-off frequency $\omega_\mathrm{cut}^\mathrm{HHG}$ scales linearly with the (peak) intensity $I_0$ of the driving pulse \cite{Krause1992,Corkum1993}, for HHG from bulk dielectrics or semiconductors $\omega_\mathrm{cut}^\mathrm{HHG}$ scales linearly with the peak field $F_0$ (or $\sqrt{I_0}$) \cite{Ghimire2014}. The processes underlying solid-state HHG have remained a matter of debate. Several simplifying models have been proposed accounting for Bloch oscillations within a single band (``intraband harmonics'') \cite{Ignatov1976,Feise1999} and non-linear interband polarization (``interband harmonics'') \cite{Schubert2014,Golde2011,Vampa2015} as sources of HHG. Most descriptions involve the semiconductor Bloch equations (SBE, \cite{Golde2008}) using input parameters on various levels of sophistication and a varying number of energy bands \cite{Hawkins2015,Wu2015}. Recently, first simulations employing time-dependent density functional theory (TDDFT, \cite{Runge1984}) have become available \cite{Otobe2016,Tancogne-Dejean2017}.

One major puzzle has remained so far unresolved: while many experiments display remarkably ``clean'' harmonic spectra with pronounced peaks near multiples of the driving frequency (odd multiples when inversion symmetry is preserved) all the way up to the cut-off frequency, corresponding simulations display a noisy spectrum lacking any clear harmonic structure over a wide range of frequencies in the ``plateau'' region above the band-gap energy. In previous theoretical works \cite{Vampa2014,Luu2015,Hohenleutner2015,Luu2016,Yu2016,You2017}, this problem was addressed by proposing remarkably short dephasing times $T_2$ in the SBEs of the order of $T_2 \approx 1$ fs or less than an optical half-cycle. While such short decoherence times yield ``cleaner'' harmonic spectra in qualitative agreement with the experiment, they raise important questions as to the ultrafast decoherence processes for electronic excitations in solids.

The point of departure of the present work is the first realistic \textit{ab-initio} multi-scale simulation of HHG by self-consistently treating the microscopic non-linear response and the mesoscopic propagation of the optical signal. Diamond can serve as a prototypical bulk dielectric for which  a full \textit{ab-initio} treatment is feasible. On the microscopic scale we employ two complementary methods: an \textit{ab-initio} TDDFT simulation and a multi-band SBE approach with input parameters from \textit{ab-initio} ground-state DFT calculations and a dense 3-dimensional Brillouin zone (BZ) sampling. Properties of the extended solid target on the mesoscopic scale enter through propagation effects on the light fields in dense matter and the inhomogeneous field distribution within the focal spot that is typically smaller than the extended target. These mesoscopic effects turn out to have a surprisingly strong influence on the resulting HHG spectra.

We use TDDFT in a real-space real-time implementation \cite{Yabana2012,Wachter2015} to simulate the electronic dynamics driven by the strong IR field $\vec{F}(t)$ employing the adiabatic local-density approximation (LDA). Alternatively, we implement the SBEs by propagating the elements of the density matrix $\rho^{\vec k}_{mn}$,
\begin{eqnarray}
\partial_t \rho^{\vec k}_{mn} & = -&i \omega^{\vec k + \vec A/c}_{mn} \rho^{\vec k}_{mn}
                             - (1-\delta_{mn})\frac{\rho^{\vec k}_{mn}}{T_2} \label{SBE} \\
     & & + i\,\vec F(t) \cdot 
      \sum_{l} \left (\vec d^{\;\vec k + \vec A/c}_{ml} \rho^{\vec k}_{ln}
      -  \vec d^{\;\vec k + \vec A/c}_{ln} \rho^{\vec k}_{ml} \right) \nonumber
\end{eqnarray}
with the transition energy $\omega^{\vec k}_{mn} = \epsilon^{\vec k}_m - \epsilon^{\vec k}_n$ and the transition dipole elements $\vec d^{\;\vec k}_{mn}$ between Houston orbitals of the valence (VB) and conduction bands (CB), $n$, $| n, \vec k + \vec A/c \rangle$ \cite{Houston1940,Krieger1986} at wave vectors $\vec k$ displaced by the time-dependent vector potential $\vec A/c$. Both $\omega^{\vec k}_{mn}$ and $\vec d^{\;\vec k}_{mn}$ are taken from ground state (GS) \textit{ab-initio} DFT serving also as initial state of the TDDFT calculation. The second term on the right-hand side describes the decoherence characterized by a dephasing time $T_2$ (a corresponding term accounting for the population relaxation with relaxation time $T_1$ is omitted because $T_1 \gg T_2$).

The crystal is irradiated by pulses polarized along the $\Gamma-\mathrm{X}$ direction having a total duration of 32~fs (corresponding to $\tau_\text{p}\approx 6.8$~fs full-width at half maximum of the peak intensity), a carrier wavelength of 800 nm [photon energy $\omega_\text{IR} = 1.55$~eV], peak intensities $I_0$ of up to $2\times 10^{13}$~W/cm$^2$, and a $\sin^6$-envelope for the field. From the time-dependent induced current densities $\vec J(t)$ derived from either methods the resulting high harmonic spectrum is calculated using Larmor's formula \cite{Jackson}
\begin{equation}
  S_{\hat n} (\omega) \propto \left| \mathfrak {Ft} \left\{ \frac d {dt} \vec J(t) \cdot \hat n \right\} \right|^2
  = \omega^2 \left| \vec J(\omega) \cdot \hat n \right|^2
  \label{eq:larmor}
\end{equation}
where $\hat n$ is the unit vector in polarization direction.

We account for the propagation through the extended system by combining the microscopic SBE solution for the current density $\vec J(t)$ of individual cells with the solution of Maxwell's equations \cite{Yabana2012}
\begin{equation}
\frac{1}{c^2}\partial^2_t \vec A(\vec r,t)-\nabla^2 \vec A(\vec r,t)= \frac{4\pi}{c} \vec J(\vec r,t)
\label{eq:maxwell}
\end{equation}
where the cells are placed on a mesoscopic grid along the propagation direction [$\vec A(\vec r,t) \rightarrow \vec A(X,t)$] with grid spacing $\Delta X=8$~nm and a crystal thickness of up to 1~$\mu$m. We use a 5-point stencil for the approximation of the second derivative in space and a standard 4-th order Runge-Kutta propagator to solve the differential equations in time. The microscopic responses to the impinging pulse at different grid points $i$, $\vec J(X_i, t)$, are thus coupled via Eq.\ \eqref{eq:maxwell}. In particular, high harmonic radiation emitted from the source term $\vec J(X_i,t)$ propagates through the crystal via Eq.\ \eqref{eq:maxwell} driving the response at the other microscopic sites thereby accounting for re-absorption of HHG within the crystal. From the electric field $\vec{F}(\omega)$ of the transmitted pulse we retrieve the harmonic spectrum via $S_{\hat{n}}(\omega) \propto \omega^2 |\vec{F}(\omega)\cdot\hat{n}|^2$.
\begin{figure}
	\includegraphics[width=0.95\columnwidth]{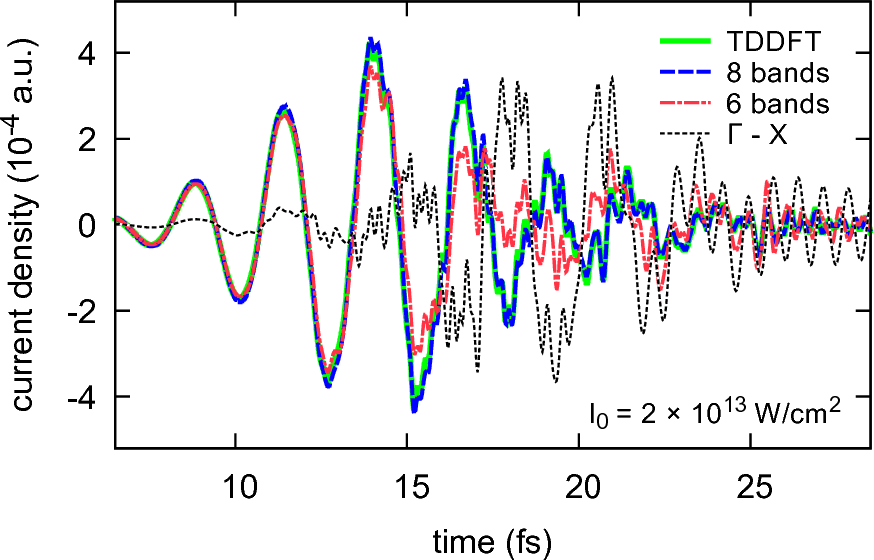}
	\caption{(Color online) Current densities induced in diamond by a 6.8~fs laser pulse with intensity $I_0=2\times 10^{13}$~W/cm$^2$ simulated using different methods: TDDFT (green solid line) and semiconductor Bloch equations (SBE) with 2 valence bands (VB) and 4 conduction bands (CB) (dash-dotted red line), 4 VBs and 4 CBs (dashed blue line) each sampled on a dense grid over the whole 3-dimensional Brillouin zone (BZ). For comparison the result for a single line in the BZ ($\Gamma - \mathrm{X}$; dotted black line; scaled) is shown. All results represent microscopic single-cell calculations neglecting mesoscopic propagation effects.}
	\label{fig1}
\end{figure}

Within a microscopic calculation employing a single cell of the periodic structure and neglecting propagation we have first verified that for moderate intensities of the driving pulse ($I_0 \sim 10^{12}$~W/cm$^2$) the SBE results for the time-dependent current density rapidly converge to that of the TDDFT prediction when consistent input for the band structure is used and dephasing is neglected ($T_2 \rightarrow \infty$). Enforcing a continuous phase evolution of the dipole matrix elements along any trajectory through the BZ, in particular near narrow avoided crossings (see Fig.\ S1) and a fine $k$-grid ($\Delta k \simeq 0.01$) have been found to be crucial prerequisites for convergence to the \textit{ab-initio} TDDFT simulation. Another important convergence parameter is the number of VBs and CBs \cite{Hawkins2015,Wu2015}. The present results (Fig.\ \ref{fig1}) show, however, that approximations including only a one-dimensional cut through the BZ and, thus, without properly sampling the full BZ as have been frequently employed \cite{Hawkins2015,Luu2015,Luu2016,Yu2016} fail in the non-linear regime.

\begin{figure}
	\includegraphics[width=\linewidth]{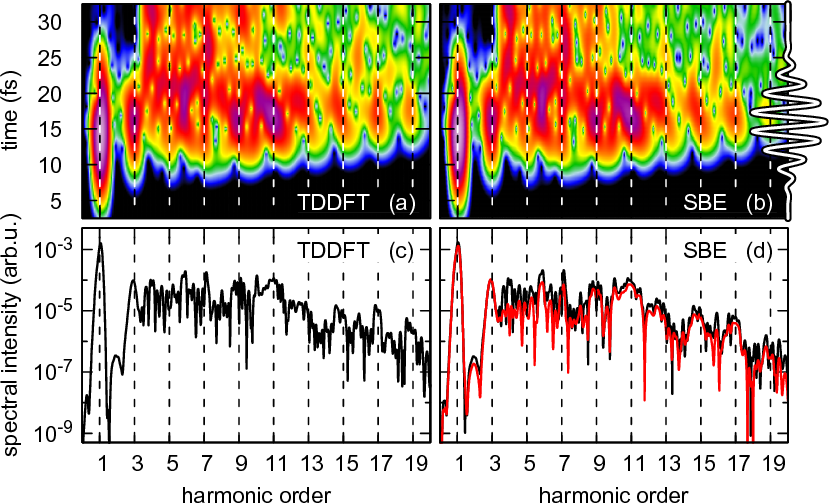}
	\caption{(Color online) Microscopic single-cell calculations of harmonic radiation emitted from diamond for $I_0=2\times 10^{13}$~W/cm$^2$, $\lambda = 800$~nm and $\tau_\text{p} = 6.8$~fs (FWHM) of the driving pulse [white solid inset in (b)]; (a) and (c) TDDFT; (b) and (d) SBEs. Upper panels (a) and (b) time-frequency analysis (Gabor transform with width $\sigma = 2$~fs; on a logarithmic color scale). Lower panels (c) and (d) resulting high-harmonic spectrum. In (d) also the spectrum including dephasing with $T_2=10$~fs (red line) is shown. The dashed vertical lines indicate the odd multiples of the carrier frequency $\omega_\mathrm{IR} = 1.55$~eV of the exciting laser pulse.}
	\label{fig2}
\end{figure}

The HHG resulting from microscopic single-cell calculations using either TDDFT or SBEs (Fig.\ \ref{fig2}) displays a noisy spectrum with strong spectral contributions at all energies above the band-gap energy ($\epsilon_\mathrm{gap}^\mathrm{LDA}\approx 5.5$ eV $\approx 3.55\,\omega_\mathrm{IR}$). The lack of clear signatures of discrete harmonics in single-cell HHG spectra of an extended periodic system can be easily understood within a semiclassical picture \cite{Vampa2015}. Within the reduced real-space zone scheme of a single cell the electron and hole wave packets driven by the pulse traverse the cell many times meeting each other at a multitude of different recombination times and with different recombination energies. Only for high frequencies near the cut-off harmonic peaks become more clearly visible as the number of contributing recombination times is strongly reduced.

The broad quasi-continuous spectrum in the plateau region (black lines in Fig.\ \ref{fig2}) is at variance with a large number of solid-state HHG experiments finding a clean harmonic spectrum \cite{Ghim2011,Schubert2014,Luu2015,Ndabashimiye2016}. A recently employed phenomenological approach to ``purify'' the spectrum has invoked including very short dephasing times of the order of a fraction of an optical cycle ($T_2\approx T_0/4$ \cite{Vampa2014}) or of about 1 fs \cite{Luu2015,Luu2016} into the microscopic description. Such short $T_2$ raise, however, conceptual questions as to the origin of such ultrafast relaxation channels for electronic excitations.

\begin{figure}
	\includegraphics[width=0.95\columnwidth]{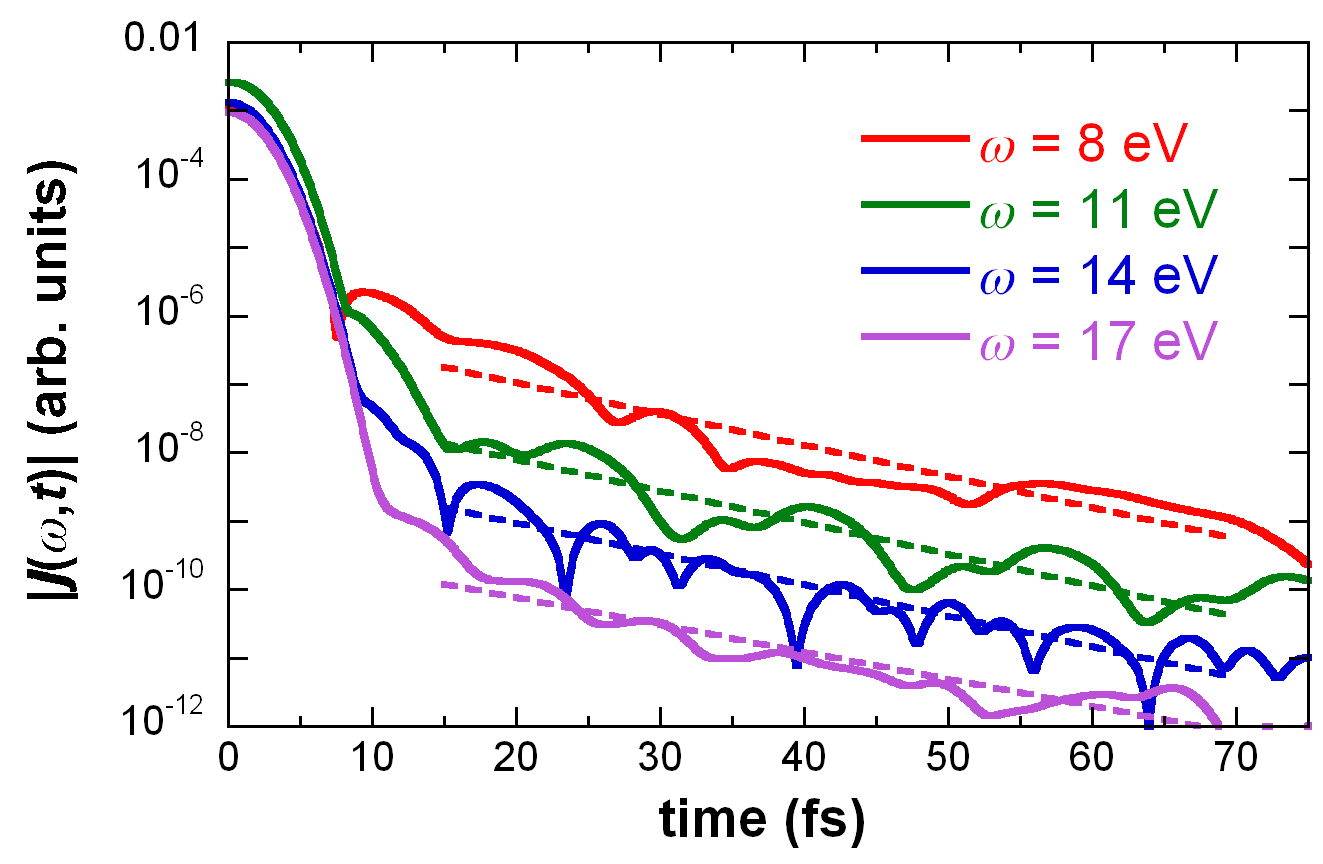}
	\caption{(Color online) Time-dependent linear response of current density to $\delta$-like (broadband) excitation at $t=0$ given by experimental optical conductivity data (\cite{Lide}, see text) evaluated at various photon energies $\omega_i$ by a Gabor transform ($\sigma = 2$~fs). The time dependent current components $J(\omega_i,t)$ exemplarily shown for $\omega_i = 8$~eV (red), 11~eV (green), 14~eV (blue), and 17~eV (purple) are fitted to exponentials $\propto e^{-t/T_2(\omega_i)}$ (dashed lines) yielding $T_2(\omega_i) \approx 10$~fs for all photon energies $\omega_i$.}
	\label{fig3}
\end{figure}

Starting point of our quantitative estimate for $T_2$ are experimental data on the optical conductivity of dielectrics. We determine the frequency dependent conductivity $\sigma(\omega)$ from the experimental complex refractive index $\sqrt{\varepsilon(\omega)}=n(\omega)+i\kappa(\omega)$ \cite{Lide} via
\begin{equation}
\sigma(\omega)=\frac{\omega}{4\pi i}[\varepsilon(\omega)-1]\, .
\label{eq:sigma}
\end{equation}
For an impulsive broad-band excitation spectrum $F(t) \propto \delta(t)$ the time-dependent induced current density follows as
\begin{equation}
J(t) = \frac{1}{2\pi}\int d\omega\, e^{-i\omega t}\sigma(\omega)F(\omega)
\label{optcur}
\end{equation}
using Ohm's law. From Eq.\ \eqref{optcur} we deduce the frequency dependent decay constant $T_2 (\omega)$ from fitting exponentials $\propto e^{-t/T_2(\omega_i)}$ to the Gabor transform of Eq. (5) at various photon energies $\omega_i$ (Fig.\ \ref{fig3}).
For frequencies above the band gap (between 8 and 17 eV) we find $T_2(\omega_i)\gtrsim 10$~fs considerably longer than previously assumed. It should be noted that this estimate is strictly valid only in the linear response regime. However, the present \textit{ab-initio} multi-scale simulation allows to extend this estimate into the non-linear regime. The non-linear extinction coefficient at the IR driving frequency $\kappa(\omega_\text{IR},I_0)$ can be directly determined from the attenuation of the driving field self-consistently propagated through the crystal. The resulting $\kappa$ is found to be strongly dependent on $T_2$ used in the simulation. Only for large $T_2$ ($> 10$~fs) corresponding to small spectral broadening we find agreement with first experimental data \cite{zz} on non-linear $\kappa(\omega_\text{IR},I_0)$ for diamond in the $10^{12}$ to $10^{13}$~W/cm$^2$ regime. Using $T_2=10$~fs as lower bound in the SBEs (Fig.\ \ref{fig2}d and SI) shows that within a microscopic single-cell calculation, the lack of pronounced high harmonics in the plateau regime persists. Therefore, inclusion of mesoscopic effects of the extended system within a multiscale treatment is crucial to describe the build-up of a clean HHG spectrum.

\begin{figure}
	\includegraphics[width=\linewidth]{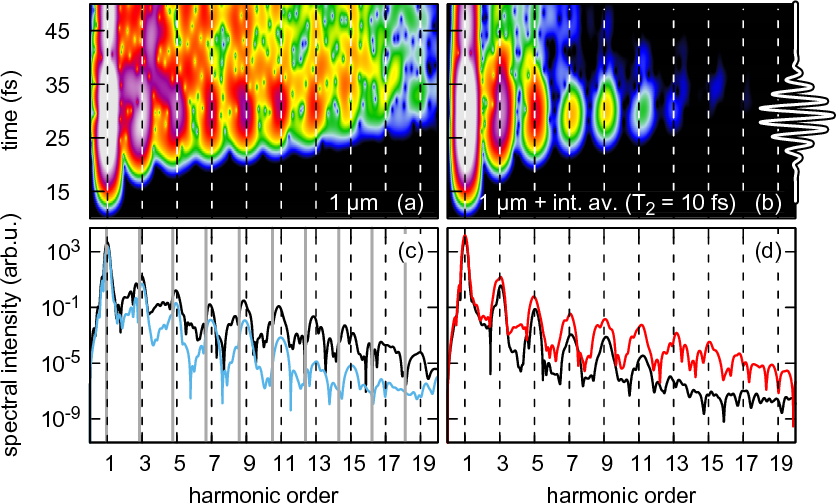}
	\caption{(Color online) Time-frequency analysis (Gabor transform with width $\sigma = 2$~fs; upper panels with a logarithmic color scale) and harmonic spectrum (lower panels) induced by a pulse with $\lambda = 800$~nm, $\tau_\text{p} = 6.8$~fs (FWHM) and incident (vacuum) peak intensity $I_\text{vac} = 2 \times 10^{13}$~W/cm$^2$ from self-consistently solving the Maxwell-Bloch equations for diamond for a layer thickness of 1~$\mu$m. (a) and (c) harmonic response in absence of dephasing ($T_2\rightarrow\infty$); (b) and (d) with $T_2=10$~fs and average over the inhomogeneous intensity distribution of the driving field. The dashed vertical lines indicate the odd multiples of the frequency $\omega_\text{t}$ of the blue-shifted transmitted pulse (see text). For comparison the odd multiples of the frequency $\omega_\mathrm{IR}$ of the incident pulse are shown as solid gray lines in (c). In (c)/(d) also the corresponding HHG spectrum with/without dephasing is shown in blue/red.}
	\label{fig4}
\end{figure}

The simulation including propagation (Eq.\ \eqref{eq:maxwell}) displays clearly visible harmonic peaks and a strongly reduced background even when dephasing is switched off ($T_2 \rightarrow \infty$ in Fig.\ \ref{fig4}a,c, layer thickness $1\mu$m). Obviously, destructive interference between a multitude of electron-hole recombination events along the propagation path at different recombination times is key to the formation of a clean harmonic spectrum with peaks at odd harmonics. It is important to note that the peak positions agree with odd multiples of the blue-shifted transmitted driving frequency $\omega_\text{t}$ rather than the incident frequency $\omega_\mathrm{IR}$ (see Fig.\ \ref{fig4}c). This predicted non-linear blue shift $\delta\omega = \omega_\text{t}-\omega_\mathrm{IR}$ in diamond is another signature of the strong non-linear response of the solid and has been, indeed, experimentally observed in different materials \cite{Kuo1992,Ghim2011}.
For the 1~$\mu$m thick crystal the directly transmitted pulse is followed by a second less intensive pulse from double internal reflection at the back and front boundaries of the slab. Due to its reduced intensity, the reflected pulse only weakly affects the total transmitted spectrum. The time delay between direct and reflected pulses ($\sim 17$ fs) allows for a direct and independent determination of the effective index of refraction of $n_\mathrm{eff}\approx 2.4$ (the real part of $\sqrt{\varepsilon(\omega)}$) in agreement with the experimental value at the carrier wavelength $\lambda=800$~nm \cite{Lide}.

Mesoscopic-scale effects include not only the propagation along the propagation direction but also the inhomogeneous field distribution within the focal spot in the transverse direction. The spot size is, typically, much smaller than the sample. Consequently, the intensity distribution of the laser pulse in the material has to be modeled adequately. The latter effect is included in this work by an average of the harmonic emission over a Gaussian profile $ I(\rho) = I_\text{max} \exp (-\rho^2/2R^2) $ of the driving laser pulse. To assess this effect independently, we calculate the resulting averaged current
\begin{equation}
\langle \vec J(t)\rangle = \frac{1}{R^2} \int_0^\infty d\rho\, \rho\, \vec J[t,I(\rho)]
\end{equation}
by a weighted sampling of the beam cross-section for an ensemble of 20 different intensities. Similar to the destructive interference due to propagation effects, also intensity averaging leads to spectra displaying more pronounced harmonic peaks with maxima at odd multiples of the driving frequency (Fig.\ S2 in SI).

Combining now the effects of destructive interference and, thus, dephasing due to propagation and transverse intensity averaging with dephasing due to microscopic decoherence processes with dephasing time $T_2=10$~fs (Fig.\ \ref{fig4}b and d) yield a well-defined harmonic spectrum in accord with experimental data. From the Gabor transform it is obvious that the effect of the finite microscopic dephasing time $T_2$ is primarily the damping of the (induced) post-pulse current $\vec J(t)$ on the time scale of a few tens of femtoseconds. It is, however, of minor importance for the formation of a clean harmonic spectrum.

In conclusion, we have presented the first \textit{ab-initio} multi-scale simulation for solid-state high harmonics self-consistently combining the microscopic non-linear response in 3 dimensions within the framework of TDDFT or multiband Bloch equations with mesoscopic propagation and source distribution effects. Irrespective of the level of the underlying description, the microscopic simulation of the non-linear response of a single cell of the periodic system of diamond alone fails to yield a well-defined harmonic spectrum. It is the spatio-temporal distribution of the emission events on the mesoscopic scale that leads to the formation of a clean high-harmonic spectrum with pronounced peaks at odd harmonics of the blue-shifted driving frequency in the medium. This process can be viewed as the solid state analog to the shaping of harmonics by propagation in extended gaseous targets due to phase matching and suppression of the contributions from long trajectories \cite{bederson1998}. Propagation and field inhomogeneity effects play, however, a much more prominent role at solid-state densities as in their absence the well-defined harmonic spectrum is replaced by a noisy quasi-continuum.
Ultrafast microscopic dephasing rates of the order of $T_2\approx1$~fs previously invoked are neither necessary nor justified for forming a well-defined harmonic spectrum. Instead, the present multi-scale simulation strongly suggests that the dephasing time should be at least one order of magnitude larger to yield the experimental optical conductivity at low intensities and the extinction coefficient in the non-linear regime of higher intensities.

The present work addressing diamond as a prototypical bulk dielectric has also wider important implications for other materials. The strong influence of propagation and field inhomogeneity effects in the dense medium on the non-linear optical response observed here is expected to be present irrespective of the details of lattice or electronic structure of the material. The blue shift of the harmonic spectrum found in the present multi-scale simulation for diamond has been, in fact, experimentally previously observed for other materials. The present findings may also contribute  to disentangling of the relative importance of intraband and interband contributions to the harmonic generation. Since interband polarization is much more effectively suppressed by previously proposed ultrashort dephasing times than coherent intraband dynamics, the contribution of interband polarization may be significantly higher than expected when realistic dephasing times are employed.
The present findings have also important implications for other ultra-fast processes, for example, probing of biomaterials by transmission  of a broadband pulse \cite{hulst2014}. Extraction of information on the presence of specific molecules will require the disentangling of the non-linear molecular response from mesoscopic light transport effects.

\acknowledgments
This work was supported by the FWF Austria (SFB-041 ViCoM, SFB-049 NextLite, doctoral college W1243, and P21141-N16), the COST Action CM1204 (XLIC), and the IMPRS-APS. It was also supported by JSPS KAKENHI Grant Numbers 16K05495, 15H03674, 26-1511, and by CREST, JST, under grant number JPMJCR16N5. Calculations were performed using the Vienna Scientific Cluster (VSC) and the supercomputer at Nagoya University through HPCI (hp160116). The authors thank Eberhard Riedle, Elisa Palacino Gonzales, Dmitry Zimin, Martin Schultze, and Nick Karpowicz for helpful discussions.

\bibliography{ms}
\end{document}


\renewcommand{\thepage}{S\arabic{page}} \setcounter{page}{1}
\renewcommand{\thesection}{\Alph{section}} \setcounter{section}{0}
\renewcommand{\thefigure}{S\arabic{figure}} \setcounter{figure}{0}
\renewcommand{\theequation}{S\arabic{equation}} \setcounter{equation}{0}
\renewcommand{\bibnumfmt}[1]{[S#1]}
\renewcommand{\citenumfont}[1]{S#1}

\centering{\large\bf Supplemental Data and Figures \\
       \textit{Ab-initio} multi-scale simulation of high-harmonic generation in solids}\\[1cm]
\justify

\FloatBarrier
\begin{figure}[h]
\includegraphics[width=0.95\columnwidth]{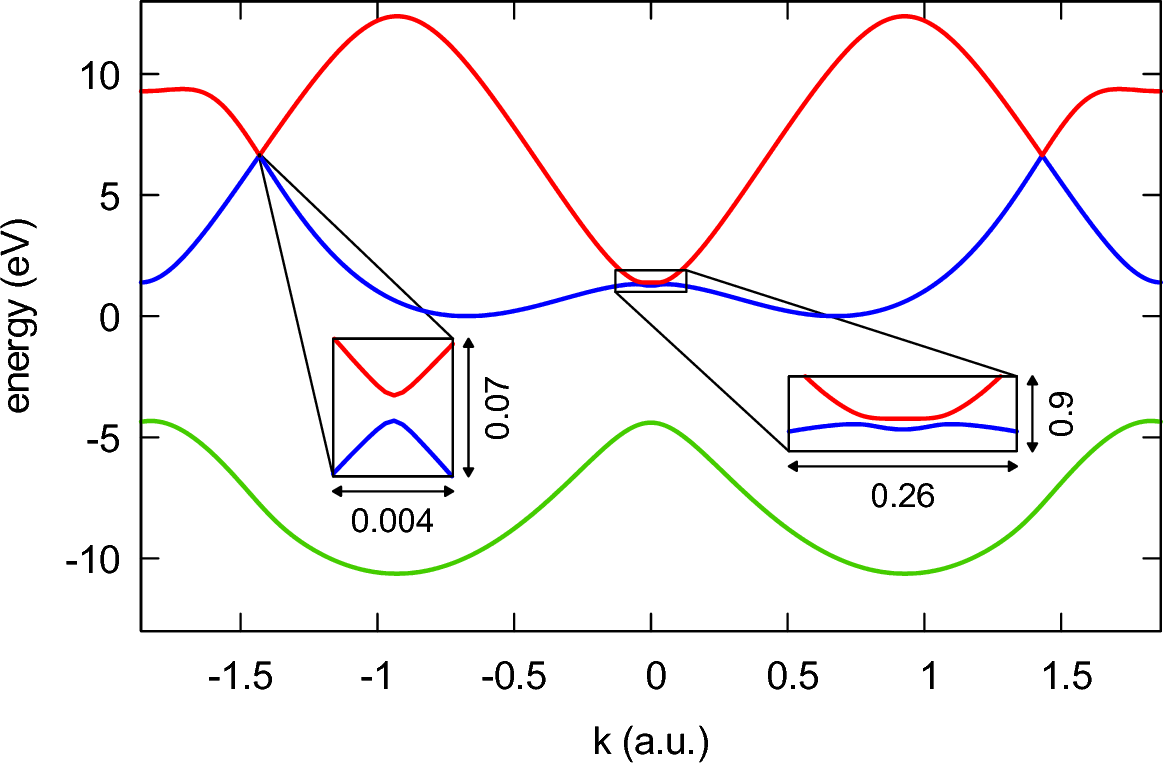}
\caption{(Color online) Cut through the Brillouin zone of diamond (calculated from DFT using LDA) along a line parallel to $\Gamma-$X [which is at $\vec k_\bot= (0,0)$] with $\vec k_\bot=(0.03,0.03)$ displaying narrow avoided crossings. The bands at the avoided crossing on the left (left inset with a total $\Delta k = 0.004$~a.u.) are separated by $\Delta E = 13$~meV. Agreement between TDDFT and SBE results crucially depends on such detailed representation of the band structure.}
\label{si_fig1}
\end{figure}

\begin{figure}[h]
\includegraphics[width=7.8cm]{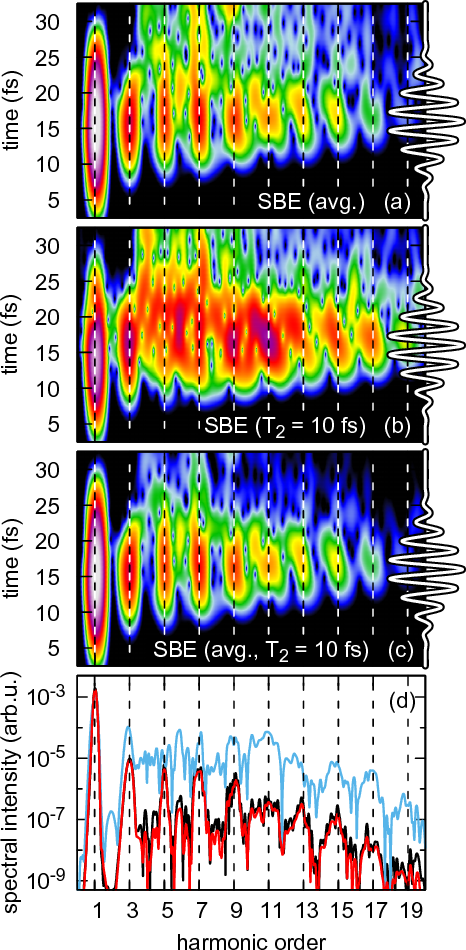}
\caption{(Color online) Time-frequency analysis (Gabor transform with width $\sigma = 2$~fs on a logarithmic color scale) of the radiation emitted from diamond; single-cell calculations for $\lambda = 800$~nm, $\tau_\text{p} = 6.8$~fs (FWHM) and $I_0 = 2 \times 10^{13}$~W/cm$^2$ (intensity in the material) of the driving laser pulse shown on the right-hand side. (a) SBE calculation as in Fig.\ 2 (main text) without dephasing ($T_2 \rightarrow \infty$) but with averaging over the transverse field distribution [Eq.\ (6) in main text], (b) SBE calculation as Fig.\ 2 (main text) without field averaging but with dephasing ($T_2=10$~fs), (c) SBE calculation with both $T_2=10$~fs and transverse field averaging, (d) resulting harmonic spectra corresponding to (a), (b), and (c) shown in black, blue, and red, respectively. The spectra are scaled to match the intensity of the first harmonic. The dashed vertical lines indicate the odd multiples of the carrier frequency of the exciting laser pulse.}
\label{si_fig2}
\end{figure}

\begin{figure}
\includegraphics[width=0.95\columnwidth]{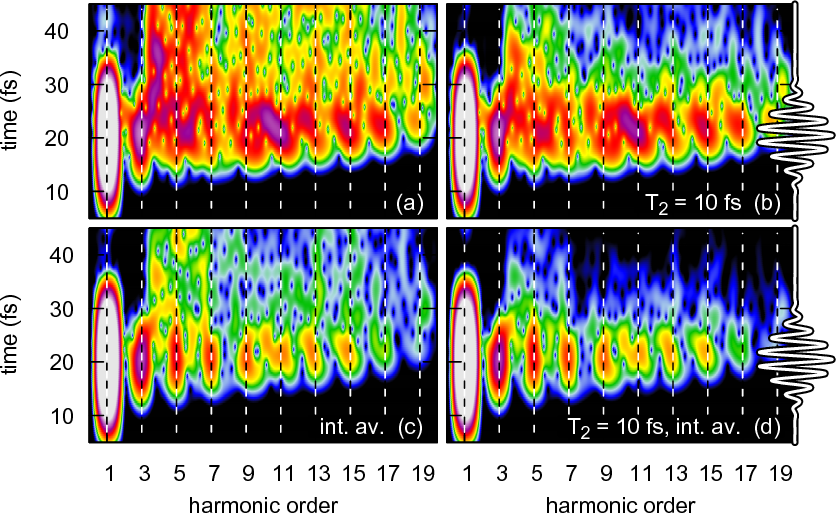}
\caption{(Color online) Time-frequency analysis (Gabor transform with width $\sigma = 2$~fs on a logarithmic color scale) of the emitted radiation induced by a laser pulse with $\lambda = 800$~nm, $\tau_\text{p} = 6.8$~fs (FWHM) and incident (vacuum) intensity $I_\text{vac} = 2 \times 10^{13}$~W/cm$^2$ (same parameters as single cell calculation in Fig.\ 2a and b) from self-consistently solving the Maxwell-Bloch equations for diamond with a thickness of $\lesssim 1$~nm (single layer) without ($T_2 \rightarrow \infty$) (a,c) and with dephasing ($T_2 = 10$~fs) (b,d). Without field averaging (a,b) and with field averaging (c,d). The dashed vertical lines indicate the odd multiples of the carrier frequency $\omega_\text{IR}$ of the incident laser pulse. Note that for $\lesssim 1$~nm thick diamond both the reduction of intensity inside the material due to reflection as well as the blue shift of the driving frequency (see main text) are negligible.}
\label{si_fig3}
\end{figure}

\begin{figure}
\includegraphics[width=0.95\columnwidth]{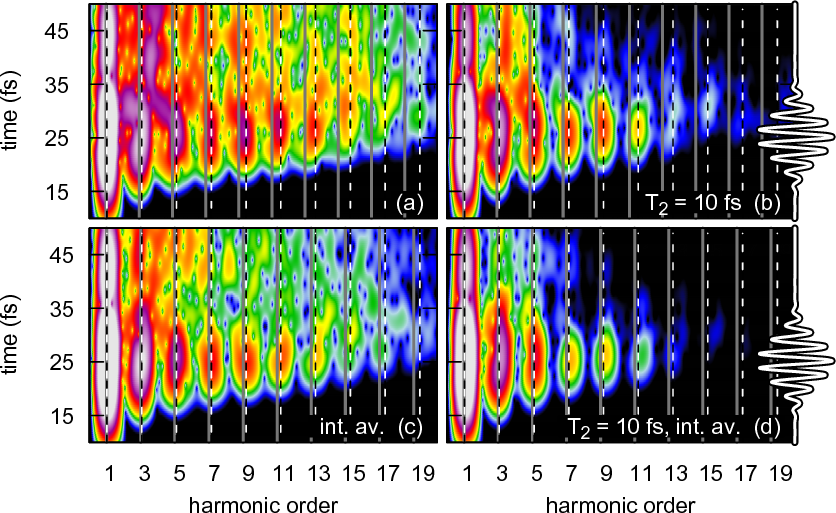}
\caption{(Color online) Same as Fig.\ \ref{si_fig3} (logarithmic color scale extended by two orders of magnitude) but for a layer thickness of 1~$\mu$m. (a,d) correspond to Fig.\ 4a, 4b in the main text and are shown for comparison, (b) with dephasing ($T_2 = 10$~fs) and (c) with intensity averaging. The dashed vertical lines indicate the odd multiples of the frequency $\omega_\text{t}$ of the blue shifted transmitted pulse (see main text). For comparison the odd multiples of the frequency $\omega_\text{IR}$ of the incident pulse are shown as solid gray lines. It is worthwhile noting that the frequency shift $\delta \omega = \omega_\text{t} - \omega_\text{IR}$ strongly depends on the incident intensity and, thus, is reduced by intensity averaging.}
\label{si_fig4}
\end{figure}
